\documentclass[twocolumn,preprintnumbers,amsmath,amssymb]{revtex4}

\def\be{\begin{equation}}
\def\ee{\end{equation}}
\def\bea{\begin{eqnarray}}
\def\eea{\end{eqnarray}}

\def\m{\mu}
\def\n{\nu}

\def\p{\partial}
\def\a{\alpha}
\def\b{\beta}
\def\t{\theta}
\def\dbar{d\hspace{-0.38em}^{-}}

\def\11{1\hspace{-0.22cm}1}
\def\22{1\hspace{-0.18cm}1}
\def\00{0\hspace{-0.20cm}0}
\def\sl{\hspace{-0.15cm}/}

\def\i{\imath}

\def\kk{\mathbf{k}}
\def\xx{\mathbf{x}}
\usepackage{graphicx}
\usepackage{dcolumn}
\usepackage{bm}
\usepackage{amsmath}
\usepackage{amssymb}
\usepackage{epsfig}
\usepackage{color}

\begin{document}

\title{Dirac Equation in the Presence of a Gravitational Background: Temporal Non Commutative Framework and Vierbein Formalism}

\author{Abolfazl\ Jafari\footnote[1]{jafari-ab@sci.sku.ac.ir}}
\affiliation{Department of Physics, Faculty of Science,
Shahrekord University, P. O. Box 115, Shahrekord, Iran}
\date{\today }

\begin{abstract}
\begin{center}
\center{\textbf{Abstract}}
\end{center}
In this paper, we study the 
non commutative quantum dynamics of the spinorial field in
the presence of a gravitational background
within DeWitt's approach. 
By employing the vierbein formalism in the temporal state of non commutative four-dimensional space time, 
we also analyze the first level correction of the solution and Green's function in terms of the Riemann curvature tensor.
The effect of the gravitational background is imposed at least up to the first order of the Riemann curvature tensor.
\end{abstract}

\pacs{03.67.Mn, 73.23.-b, 74.45.+c, 74.78.Na}

\maketitle

\noindent {\footnotesize Keywords: Dirac Theory, Non Commutative Coordinates, Vierbein Field, Gravitational Background}

\section{Introduction}
Interestingly, the interaction between matter and gravitational waves at the quantum level
has not been explored in a large extent. 
In the current literature two different approaches can be recognized, the one by Weber and the second by DeWitt.
The first approach is rewriting quantum mechanics on a curved background
space time exclusively.
But the second way introduced by Weber as he was concerned with the response of matters to the presence of gravitational effects, and deals solely with the interaction of gravitational waves with matter.
Quite remarkable,
these two approaches happen to not produce exactly equivalent results
and also any results that will be found by one of them can not obtained from another
way \cite{DeWitt, speliotopoulos, saha}.

In this work we restrict ourselves to the weak gravitational background limit, in which
a metric description of space time is valid, and the Dirac equation generalized
to curved space time governs the dynamics of system. 
At available approximation,
we assume that the considered system is in free fall along a geodesic during the
time required for a physical event.
The typical treatment of the problem found in the literature consists of
writing the Dirac equation in curved space time and deriving correction terms to the flat
space time in perturbation theory. 
The relativistic regime of this approach yields a non commutative version of 
Dirac equation containing an effective classical geodesic deviation terms
which describe the typical interaction of the physical systems with the gravitational field.

From the Einstein principle of equivalence, local gravitational effects
are not present in a neighborhood of the space time origin of a locally inertial
frame of reference.
According to which, special relativity will be available in an free falling laboratory.
Indeed, can be found a local inertial frame for each point interior of space time $\rho_0$ at least up to
the first order of Riemann's curvature tensor.
As viewed in such frames, free particles must move along straight lines, at least locally.
Accordingly, there is a flat relevant tangent space time of the gravitational background. 
Linear gravitational waves like those coming from
the palpitating stars, are small fluctuations in metric, given by 
$h_{\m\n}(t,z)$ which propagates in the $z$-direction and is at least time and direction dependent \cite{maggirro}.
In the present work, we assume the gravitational background which is only time dependent $h_{\m\n}(t)$.
In this case, following the common approach, we decompose the metric tensor as:
\begin{align}\label{lgw-metric}
ds^2=\eta_{\m\n}dx^\m dx^\n+h_{ij}(t)dx^i dx^j,
\end{align}
where Latin indices run from 1 to 3 and Greek indices take on the values 0,1,2,3.
In which $h_{ij}$'s are regarded as very small time dependent functions.
The metric of the tangent space time is induced by the essential space time.
The induced metric is given by Eq.(\ref{lgw-metric}) and
the rising of the Lorentz indices which appear in the tangent space time are done with Eq.(\ref{lgw-metric}) \cite{parker00}.
By choosing the transverse and traceless gauge
(TT-gauge) of the gravitational background, we have  $h_{0\m}=0$. 
Also by denoting $\mathfrak{D}^\m$ as the covariant derivative, we have
$\mathfrak{D}^\m h_{\m\n}=0$.
By these tools, the gravitational background contains only two physical polarizations.
The form
\begin{align}
h_{\m\n}=\left(
\begin{array}{cccc}
0 & 0 & 0 & 0 \\
0 & 0 & h_{12} & 0 \\
0 & h_{21} & 0 & 0 \\
0 & 0 & 0 & 0
\end{array}
\right),
\end{align}
by setting $h_{12}=h_{21}$ as $"\times"$ polarization
are familiar \cite{speliotopoulos, weber, misner, inverno, ohanian}.
In the TT-gauge, the non vanishing components of the affine connections are: 
\begin{align}\label{affine}
\Gamma^0_{12}=\Gamma^1_{02}=\Gamma^2_{10}=-\frac{\dot{h}_{12}}{2},
\end{align}
for $\times$ polarization.
Also, only non zero components of the Riemann's curvature tensors are $R^0_{jk0}=\p_0\Gamma^0_{jk}=-\frac{1}{2}\dot{\dot{h}}_{jk}$ \cite{speliotopoulos, saha}.
From the perturbed metric, one finds that for the $"\times"$ polarization the only non-zero Riemann's element is $R^0_{120}$, where $"t"$ stands for the coordinate time of the proper detector frame
such as proper time which appearing in the disturbed metric to the
first order of the perturbation. 

There are arguments suggesting that the structure of space time
is drastically changed at very short distances such as scales relevant to quantum gravity.
One mathematical
candidates for the framework to study these scales happens to be the non commutative 
spaces, among which is the one known as canonical non commutativity, by which space time coordinates satisfy the following commutator 
\cite{connes, szabo, witten, nekrasov}
\begin{align}
\label{1}
[x^{\m},x^{\n}]_\star=i\theta^{\mu\nu},
\end{align}
\noindent where $[f,g]_\star=f(x)\star g(x)-g(x)\star f(x)$
and $\theta^{\mu\nu}$ is a constant, real and antisymmetric tensor.
The formulation of theories on non commutative spaces follows a very simple recipe, namely,
replacing the ordinary products between quantities by a new one, the so-called
$\star$-product \cite{szabo, nekrasov, wess}
\begin{align}\label{MoyalWeylmap}
f(x)\star g(x)=f(x) \,{\rm e}^{\frac{i}{2}\frac{\p}{\p_{x^\mu}}\theta^{\mu\nu}\frac{\p}{\p_{y^\nu}}}\ g(y)\mid_{y=x}.
\end{align}
Interestingly, the upper-bounds on various non commutative 
parameters $\t^{ij}$ happen to be close to the scales relevant to the quantum gravity effects
\cite{who3, no1, no2, no3}. 
In particular, the upperbounds around
$10^{-2}\,$TeV \cite{who1, who2, no1, no2, no3}, corresponding in units $\hbar=c=1$ to length $10^{-18}\textrm{-}10^{-20}\,$m, are quite comparable with the orders relevant in interaction of gravitational waves with present detectors \cite{speliotopoulos, saha, who1, who2, jab1, jab2, jaf4}.
Much attention has also been paid to quantum field theories on non commutative space time, in particular
the non commutative version of the Dirac theory as well as non commutative linear gravitational waves \cite{speliotopoulos, saha, jaf4}.
In contrast to spatial non commutative quantum field theory, we want to study the non commutative Dirac theory in the temporal mode
($[x^\m,x^\n]_\star=\imath\delta^{\n0}\theta^{\m\n}$)
includes a direct interaction term which is constructed by a gravitational effect.
Also, we know that the non commutativity can be expanded in terms of the Riemann curvature tensor (affine connection coefficients): $\t^{\m\n}(x)=\t^{\m\n}+\t^{\m\a}\Gamma^\b_{\a\b}x^\n+...$ therefore, we restrict ourselves to the first order of $\t\Gamma$ limit as a new expansion parameter.
It happens when we want to follow the simple recipe, that is
replacing the ordinary products between quantities by $\star$-product.

The purpose of the present work is first to derive a perturbed Dirac theorem to the gravitational background limit based on the non commutative coordinates within DeWitt's framework, and
second to develop a set of the tetrad fields based on a given algorithm.
In particular, the result is obtained by the rewriting of the motion equations
for a particle (antiparticle) interacting with the gravitational background.

As the final comment in this part, here we only consider
the interaction between the Dirac field and gravitational background.
Since, the variation magnitude of the Riemann tensors is small
so the non commutative version of Eq.(\ref{lgw-metric}) and Eq.(\ref{affine}) is yet available in terms of the \textit{Riemann coordinates} without any change in expression \cite{weber, misner, inverno, ohanian}.

\section{Vierbein Field and Generalized Dirac Theorem}
In flat space time, the Dirac theorem has the Lagrangian density: 
\begin{align}\label{free-dirac-lagrangian}
\mathcal{L}=\bar{\psi}(\imath\gamma^\m\p_\m-m\11)\psi,
\end{align}
where for four-dimensional Minkowski space time, four $\gamma^\m$-block form matrices satisfy
the Lorentz algebras and $"\11"$ is a unit operator.
The Chiral representation is an especially convenient choice, although there are many different representations.
However, in this article, we use the Chiral representation in which, $\gamma^0$ is not diagonal \cite{peskin, kleinert}.
The Chiral representation is: 
\begin{align}
\gamma^i=\begin{pmatrix}\00 &\sigma^i \\ -\sigma^i & \00 \end{pmatrix},
\gamma^0=\begin{pmatrix}\00 &\11 \\ \11 & \00 \end{pmatrix}.
\end{align}
Suppose that, we have a set of two $2*2$ matrices; $\11$ and $\00$ correspond to:
\begin{align}
\11=\begin{pmatrix}1 &0 \\ 0 & 1\end{pmatrix},\quad
\00=\begin{pmatrix}0 &0 \\ 0 & 0\end{pmatrix}.
\end{align}
We thus need only derive one explicit generalization of the Dirac theorem.
One of the ways of generalizing the Dirac theorem is employing vierbein fields.
The vierbein fields are the set of local subjects which can provide $\tilde{\gamma}^\m(x)$ for a general space time 
in terms flat space time $\gamma^\m$ matrices. 
In this work, $a^\m_{\ \a}(x)$ stands for vierbein fields.
Generally, vierbein fields $\tilde{\gamma}_\star(x)$ are functions of non commutative coordinates.
This means that the motion equation of $\tilde{\gamma}_\star(x)$ will also be different from $\tilde{\gamma}(x)$.
But in this paper, $\tilde{\gamma}_\star(x)$ are only time dependent
and satisfy the same equation of motion of $\tilde{\gamma}(x)$.
The dynamics equation of vierbein fields consists of the spinorial affine connections $\Gamma_\m(x)$.
These
are matrices defined by the vanishing of the covariant derivative of $\tilde{\gamma}^\m(x)$ matrices:
\begin{align}
\tilde{\gamma}_{\n;\m}(x)=
\p_\m\tilde{\gamma}_\n(x)-\Gamma^\lambda_{\m\n}(x)\tilde{\gamma}_{\lambda}(x)+[\tilde{\gamma}_{\n}(x),\ \Gamma_\m(x)]=0.
\end{align}
So, the covariant derivative is as follows:
\begin{align}
A_{;\m}=\mathfrak{D}_\m A_\a=
\p_\m A_\a-\Gamma^\lambda_{\m\a}A_\lambda
+[\tilde{\gamma}_\m\ ,\ A_\a],
\end{align}
the effect of which on a spinor field $\psi$, is:
$$\mathfrak{D}_\m\psi=\p_\m\psi-\Gamma_\m\psi.$$
Thus, the Lagrangian of Eq.(\ref{free-dirac-lagrangian}) generalizes to the relevant action:
\begin{align}\label{generalized-dirac-action}
\mathcal{S}=\int\ d^4x\ \sqrt{-g}\ \bar{\psi}(\imath\tilde{\gamma}^\m(x)\mathfrak{D}_\m-m\11)\psi,
\end{align}
where $"g"$ stands for determinant of the metric in which up to first order of the Riemann curvature tensors $"g"=-1+0(h^2)$.

\section{Vierbein Field-}

The $\tilde{\gamma}(x)$-matrices are space time dependent functions and obey the commutation relation:
$\{\tilde{\gamma}^\a(x),\tilde{\gamma}^\b(x)\}=2g^{\a\b}(x)\11$.
An explicit representation of the $\tilde{\gamma}^\m(x)$-matrices in terms of flat space time $\gamma^\m$-matrices is obtained by introducing 
$x$ dependent $a^{\a}_{\ \m}(x)$ vector fields.
They may be called tetrad fields which constitute four
different vector fields and must be able to produce the metric of space time.
Under transformations of the coordinates, indices $\m$, $\n$, are regarded as tensor
indices, while indices $\a$, $\b$ act merely as labels. 
In addition to the covariance under general coordinate
transformation acting on the space time indices $\m$, $\n$ the formalism is covariant
under the Lorentz transformations applied to the vierbein indices $\a$, $\b$.
Vierbein indices are lowered with $\eta_{\a\b}$, while space time indices are lowered with the metric $g_{\m\n}$.
A basic relation of the vierbein field is:
\begin{align}\label{vierbein-fields}
g_{\m\n}(x)=a^{\a}_{\ \m}(x)\eta_{\a\b}a^{\b}_{\ \n}(x),
\end{align}
in which under coordinate transformation, indices $\m,\n,..$are regarded as tensor indices,
while the remaining indices act merely as labels. 
Due to Ref.\cite{parker, birrell}, we can write 
\begin{align}\label{new-gamma}
\tilde{\gamma}^{\m}(x)=\gamma^\a a^{\ \m}_{\a}(x).
\end{align}
It is possible to calculate a set of vierbein fields for a special case of the gravitational background according to
the Ref.\cite{parker00}.
By taking account of the fact that:
\begin{align}\label{amiu2}
g_{\m\n}=\left(\begin{array}{cccc}
                 1 & 0 & 0 & 0  \\
                 0 & -1 & h & 0  \\
                 0 & h & -1 & 0  \\
                 0 & 0 & 0 & -1
               \end{array}
\right),
\end{align}
and using the proccess \cite{parker00}, 
we can determine explicitly the components of 
$a_{\ \m}^\a$
by the following formulas:
\begin{align}\label{amiu1}
a_{\ 0}^{\a}=\delta^\a_0,
\cr
a_{\ i}^\a=\delta^\a_i+\int^t\int^{\acute{t}}R^\a_{\ 0i0}(\acute{\acute{t}}) d\acute{\acute{t}}d\acute{t}.
\end{align}
we find from Eq.(\ref{amiu1}) and Eq.(\ref{amiu2}) that:
\begin{align}
a_{\ \m}^{\a}=\left(
              \begin{array}{cccc}
                1 & 0 & 0 & 0 \\
                0 & 1 & -\frac{h}{2} & 0 \\
                0 & -\frac{h}{2} & 1  & 0 \\
                0 & 0 & 0  & 1
              \end{array}
            \right).
\end{align}
We also have $a^{\a\lambda}=g^{\lambda\n}a^\a_{\ \n}$, 
$a^{\a\m}=a_{\ \n}^{\a}g^{\n\m}=\eta^{\a\b}a_\b^{\ \m}$, and $a_{\a}^{\ \m}=\eta_{\a\b}a_{\ \n}^{\b}g^{\n\m}$, therefore
\begin{align}\label{amiu1_02}
a^{\ \m}_{\a}=\left(
              \begin{array}{cccc}
                1 & 0 & 0 & 0 \\
                0 & 1 & \frac{h}{2} & 0 \\
                0 & \frac{h}{2} & 1  & 0 \\
                0 & 0 & 0  & 1
              \end{array}
            \right),
\end{align}
and hence
\begin{align}\label{amiu1_01}
a^{\a\m}=\left(
              \begin{array}{cccc}
                1 & 0 & 0 & 0 \\
                0 & -1 & -\frac{h}{2} & 0 \\
                0 & -\frac{h}{2} & -1  & 0 \\
                0 & 0 & 0  & -1
              \end{array}
            \right).
\end{align}
Continuing the evaluation of $\tilde{\gamma}^\m(x)$, we define: 
\begin{align}
\tilde{\gamma}^0=\gamma^0,\ \ \tilde{\gamma}^1=\gamma^1+\frac{1}{2}h\gamma^2,\ \ \tilde{\gamma}^2=\gamma^2+\frac{1}{2}h\gamma^1, 
\cr
\tilde{\gamma}^3=\gamma^3,
\end{align}
$\tilde{\gamma}^\m(x)$ are defined as before by Eq.(\ref{new-gamma}).
Now, according to Ref.\cite{parker, birrell}, the spinorial affine connection has the solution:
\begin{align}\label{spinorial-affine-connection-sol1}
\Gamma_\m=
-\frac{1}{4}\gamma_\a\gamma_\b a^{\a\lambda}\mathfrak{D}_\m a^{\b}_{\ \lambda},
\end{align}
we evaluate the spinorial affine connection by substituting 
$a^{\b}_{\ \lambda}$ and $a^{\a\lambda}$,
\begin{align}
\Gamma_\m=-\frac{1}{4}\tilde{\gamma}^\n\tilde{\gamma}_{\n;\m},
\end{align}
therefore, we find
\begin{align}\label{spinorial connection}
\Gamma_\m=0.
\end{align}
Define $\sigma^\m=(\sigma^0,\vec{\sigma})$ and $\bar{\sigma}^\m=(\sigma^0,-\vec{\sigma})$,
so that
\begin{align}\label{covariant-generalized-derivative}
\imath\tilde{\gamma}^\m\p_\m=\left(
              \begin{array}{cc}
                \00 & \i\sigma^\m\p_\m-\frac{\i}{2} h\Sigma^i\p_i  \\
                \i\bar{\sigma}^\m\p_\m-\frac{\i}{2} h\bar{\Sigma}^i \p_i & \00
                \end{array}\right),
\end{align}
where $\mathbf{\Sigma}=(\sigma^2,\sigma^1,0)$ and $\bar{\mathbf{\Sigma}}=(\bar{\sigma}^2,,\bar{\sigma}^1,0)$.
Finally, we see that the symmetrical structures hold in this presentation.

Based on the previous section, let us impose a non commutativity on the framework:
\begin{align}\label{new-framework}
[x^\m,x^\n]=\i\t^{\m\n}\delta^{0\m}.
\end{align}
Substituting Eq.(\ref{new-framework}) into Eq.(\ref{covariant-generalized-derivative})
From which we can present a Dirac theorem in the presence of the gravitational background in the temporal state of non commutativity.
It is enough
to replace the product of fields in the commutative actions with the star product. 
Along our general recipe, the action proposed for the non commutative Dirac
theory is
\begin{align}\label{perturbed-dirac-field}
\mathcal{S}=\int\ d^4 x\sqrt{-g}\ \bar{\Psi}\star\mathbf{N}(h)\star\Psi,
\end{align}
where
\begin{align}
\mathbf{N}(h)=\left(
              \begin{array}{cc}
                -m\11 & \i\sigma^\m\p_\m-\frac{\i}{2} h\Sigma^i \p_i  \\
                \i\bar{\sigma}^\m\p_\m-\frac{\i}{2} h\bar{\Sigma}^i \p_i & -m\11
                \end{array}\right).
\end{align}
Eq.(\ref{perturbed-dirac-field}) does not affect products involving
the metric tensor. 
Because we do not research of the
dynamic of the space time geometry.
Keeping only the terms up to first order in $"h"$ in Eq.(\ref{perturbed-dirac-field}) and due to rotation rule, we simply have:
\begin{align}\label{perturbed-dirac-equation3}
\mathcal{S}=\int\ d^4x\ \bar{\psi}(\mathbf{N}(h)\star\psi),
\end{align}
Eq.(\ref{perturbed-dirac-equation3}) can also be rewritten in the form:
\begin{align}\label{perturbed-dirac-field2}
\mathcal{S}=
\cr
\int d^4 x \bar{\Psi}(\mathbf{N}(h)\Psi+\frac{\i}{2}\t^{0i}\dot{\mathbf{N}}(h)\Psi_{,i}
-\frac{1}{4}\t^{0i}\t^{0j}\ddot{\mathbf{N}}(h)\Psi_{,ij}+...)
\cr
=\int d^4 y (\bar{\Psi}\mathbf{M}(h,\t)\Psi).
\end{align}
But, we have to calculate the perturbative contribution to lowest order of the Riemann curvature tensor,
so we can ignore higher order terms of $"h"$ and $"\t"$ in Eq.(\ref{perturbed-dirac-field2}).
Expanding the Lagrangian of Eq.(\ref{perturbed-dirac-field}) to the first order of the Riemann curvature tensor will yield the equation of motion satisfied by spinor fields:
\begin{align}\label{uone1}
(\mathbf{M}_0+\mathbf{M}_1(h,\t))\Psi(k,x)=0,
\end{align}
where 
\begin{align} \label{m01}
\mathbf{M}_0=
\left(
              \begin{array}{cc}
                -m\11 & \i\sigma^\m\p_\m \\
                \i\bar{\sigma}^\m\p_\m & -m\11
                \end{array}\right),
                \cr
\mathbf{M}_1(h,\t)=                
                -\frac{1}{2}\i\left(
              \begin{array}{cc}
                \00 & \mathbf{T}\Sigma^j \p_j  \\
                \mathbf{T}\bar{\Sigma}^j \p_j & \00
                \end{array}\right),
\end{align}
and $\mathbf{T}=h+\frac{\i}{2}\t^{0i} \dot{h}\p_i$.

\section{The Solution}
In this section, we attempt to present a solution for the perturbed Dirac theorem.
Since, Dirac field obeys the Klein-Gordon equation,
then it must be written as a linear combination of plane waves in addition correction terms.
Thus, the Dirac solution is as follows:
\begin{align}\label{sefr}
\Psi^s_{f}(k,x)=\left(
              \begin{array}{c}
                u_{0}^s(k)e^{-\imath k\cdot x} \\
                v_{0}^s(k)e^{+\imath k\cdot x}
                \end{array}\right),
\end{align}
where, $k^2=m^2$ and $"s"$ denotes the particle which is spin-up or -down along the $z$-direction.
$u^s_{0}(k)$ with $k_0>0$ whereby the solution has positive frequency (or $v^s_{0}(k)$ with $k_0<0$ whereby the solution has negative frequency), as a column vector must obey:
\begin{align}\label{uone7}
(\gamma^\m k_\m-m\11)u_{0}^s(k)=0.
\end{align}
The Euler-Lagrange equation for $v_0^s(k)$ gives the same
equation, in hermitian conjugate form:
$(\gamma^\m k_\m+m\11)v_{0}^s(k)=0.$
We know that two linearly independent solutions of free fields equations are realized:
\begin{align}
u_0^s(k)=\left(
              \begin{array}{c}
                \sqrt{k\cdot\sigma}\ \xi^s \\
                \sqrt{k\cdot\bar{\sigma}}\ \xi^s
                \end{array}\right),
\cr                
v_0^s(k)=\left(
              \begin{array}{c}
                \sqrt{k\cdot\sigma}\ \zeta^s \\
                -\sqrt{k\cdot\bar{\sigma}}\ \zeta^s
                \end{array}\right),
\end{align}
here $"s"$ takes on the values $\pm$.
$\xi$ and $\zeta$ are basis of two-components of spinors and orthogonal to each other.
$"s"$ is also useful to distinguish the right and left handed particles correspondence to $s=+,-$ along the $z-$direction \cite{peskin, kleinert, parker00}.
We also have the premise that $\mid\pm>$ are eigenvectors of the matrix $\sigma^3$. 
However, in order for Eq.(\ref{uone1}) to has a non-zero solution, the determinant of the 
matrix of coefficients Eq.(\ref{m01}) must be zero. 
We can find the the correct energy-momentum relation by making the replacement 
$"\mathbf{\p}\rightarrow\pm\i k"$ in Eq.(\ref{m01}).
The determinant be zero gives the following expression for the energy of particles with positive and negative frequency:
\begin{align}\label{deretrminant of energy}
E^\mathrm{pos}=\sqrt{m^2+\mathbf{k}\cdot\mathbf{k}-k_1k_2(h-\frac{1}{2}\t^{0i}\dot{h}k_i)},
\cr
E^\mathrm{neg}=\sqrt{m^2+\mathbf{k}\cdot\mathbf{k}-k_1k_2(h+\frac{1}{2}\t^{0i}\dot{h}k_i)}.
\end{align}
Thus, it is interesting that the non commutativity 
correction already makes a distinction between the solutions with positive and negative frequency.
Now, if we assume that $u_0(k)$ satisfies the Euler-Lagrange equation of motion for the free case, then it follows that $u^s(k)-u^s_0(k)=u^s_1(k)$ can be regarded as small. 
From which we have
\begin{align}\label{u1}
u^s_1(k)=-\frac{1}{\mathbf{M}_0}\mathbf{M}_1(h,\t)\mid_{_{\p_i\rightarrow\i k_i}}u^s_0(k),
\cr
=-\frac{\mathbf{M}_0+2m\11}{k^2-m^2}\mathbf{M}_1(h,\t)\mid_{_{\p_i\rightarrow\i k_i}}u^s_0(k),
\end{align}
but, it can be seen that 
\begin{align}\label{m1m0com}
\{\mathbf{M}_0,\mathbf{M}_1(h,\t)\}\mid_{_{\p_i\rightarrow\i k_i}}=
\cr
-2(h-\frac{1}{2}\t^{0i} \dot{h}k_i)k_1k_2\11-2m\mathbf{M}_1(h,\t).
\end{align}
Substituting Eq.(\ref{m1m0com}) into Eq.(\ref{u1}) we find,
\begin{align}\label{answer}
u^s_1(k)=2\frac{(h-\frac{1}{2}\t^{0i} \dot{h}k_i)k_1k_2}{k^2-m^2}u^s_0(k).
\end{align}
Certainly, there is no $H_D=-\i\mathbb{\a}\cdot\mathbb{\nabla}_q+m\b$ as a time independent operator,
because this may be questioned by $h(t)$-gravitational background.
This means that there is no longer a way to find a boost to reach the rest frame.
So, the rest frame can not occur according to the argument of the interaction.
We should emphasize that Eq.(\ref{answer}) is only valid within the rest frame.
Therefore, we must add a correction term $\mathfrak{Z}(k,h,\t)$ to Eq.(\ref{answer}): 
\begin{align}\label{answer0}
u^s_1(k)=(2\frac{(h-\frac{1}{2}\t^{0i} \dot{h}k_i)k_1k_2}{k^2-m^2}+\mathfrak{Z}(k,h,\t))u^s_0(k).
\end{align}
Thus, we finally arrive at an expression for the solution with positive frequency:
\begin{align}\label{answer00}
u^s(k)=(1+2\frac{(h-\frac{1}{2}\t^{0i} \dot{h}k_i)k_1k_2}{k^2-m^2}+\mathfrak{Z}(k,h,\t))u^s_0(k)
\cr
=:\b(k,h,\t)u^s_0(k).
\end{align}
It can be seen that $u^\dagger(k)=\b(k,h,\t)u_0^\dagger(k)$, thus $\bar{u}(k)=\b(k,h,\t)\bar{u}_0(k)$
Also, we want to have 
\begin{align}
u^{s\dagger}(k)u^r(k)=2E^\mathrm{pos}_0\b^2(k,h,\t)\delta^{rs}=:2E^\mathrm{pos}\delta^{rs},
\end{align}
namely, the energy of these particles becomes:
\begin{align}\label{corrected-energy}
E^\mathrm{pos}=E^\mathrm{pos}_0\b^2(k,h,\t).
\end{align}
Comparing Eq.(\ref{corrected-energy}) with Eq.(\ref{deretrminant of energy}), we find that
\begin{align}
\mathfrak{Z}(k,h,\t)=-2\frac{(h-\frac{1}{2}\t^{0i}\dot{h}k_i)k_1k_2}{k^2-m^2}-\frac{(h-\frac{1}{2}\t^{0i} \dot{h}k_i)k_1k_2}{4E^2_0},
\end{align}
by substituting this result, Eq.(\ref{answer00}) becomes: 
\begin{align}\label{answer0000}
u^s(k)=(1-\frac{(h-\frac{1}{2}\t^{0i} \dot{h}k_i)k_1k_2}{4E^2_0})u^s_0(k)=\b(k,h,\t)u^s_0(k).
\end{align}
Now, $\b(k,h,\t)$ is the correction factor of mass.
Thus,
\begin{align}\label{correction field mass}
m^\mathrm{pos}=m^\mathrm{pos}_0\b^2(k,h,\t).
\end{align}
as a new field mass will lead us to:
\begin{align}
\bar{u}^s(k)u^r(k)=-2m^\mathrm{pos}\delta^{rs}.
\end{align}
Similarly, for obtaining the solution with negative frequency, it is sufficient to replace $k_\m\rightarrow-k_\m$ in Eq.(\ref{answer00}). 
So, the solution with negative frequency will be:
\begin{align}\label{answer000}
v^s(k)=\b(-k,h,\t)v^s_0(k).
\end{align}
with
\begin{align}
v^{s\dagger}(k)v^r(k)=2E^\mathrm{neg}\delta^{rs}, 
\cr
\bar{v}^s(k)v^r(k)=-2m^\mathrm{neg}\delta^{rs},
\end{align}
with $m^\mathrm{neg}=m_0\b^2(-k,h,\t)$.
We mention that, the solutions with positive and negative frequency are treated in a different way. 
We shall use distinct notations for the two types of the particles:
solutions with positive and negative frequency.
Therefore, for a negative-frequency solution, $\b^2(-k,h,\t)$ is the correction factor to the energy and mass.

To complete the derivation of the solution, a summation over $"\mathbf{k}"$ is understood,
\begin{align}
\psi(x)=
\cr
\int \dbar{}^3k  \Sigma_s(\frac{1}{\sqrt{2q^+_0}}a^s_\kk u^{s}(\kk)+\frac{1}{\sqrt{2q^-_0}}b^{s}_{-\kk} v^{s}(-\kk))
{\rm e}^{-\i \kk\cdot\xx}
\cr
=\int \dbar{}^3k \frac{1}{\sqrt{2k_0}} \Sigma_s\Big{(}a^s_\kk u^{s}_0(\kk)
+b^{s}_{-\kk} v^{s}_0(-\kk)\Big{)}
{\rm e}^{-\i \kk\cdot\xx},
\end{align} 
with $q^\pm_0=k_0\b^2(\pm k,h,\t)$. 
Where $\dbar{}^3 k=\frac{d^3k}{(2\pi)^3}$, $b^{s}$ and $a^s$ are operators coefficients.
So, from these considerations and for the temporal state of non commutativity, we can conclude that 
all of the completeness relations are exactly the same as in the unperturbed Dirac theorem.
Nevertheless, we must admit that of change in the many conceptions of perturbed Dirac theorem, 
such as destroying the local conserved currents $j^\m=\bar{\Psi}\star\tilde{\gamma}^\m(x)\star\Psi$.
Because,
\begin{align}
\p_\m j^\m=
\cr
+\i\bar{\psi}[\gamma^\m,M_1]\psi
+\frac{1}{2}\t^{0i}\bar{\psi}[M_1\p_i,\gamma^\m]\psi,
\end{align}
that is the local currents are no longer exactly conserved. 
Let us now investigate Green's function of this perturbed theory.
Eq.(\ref{uone1}), which defines Green's function can be rewritten as
\begin{align}\label{gf0}
(\i\gamma^\m\p_\m-m\11+\mathbf{M}_1(h,\t))\hat{G}(x,y)=\i\delta{(x-y)},
\end{align}
the change in the momentum representation of Eq.(\ref{gf0}) ($\hat{G}(x,y)=\int \dbar{}^4 k {\rm e}^{-\i k\cdot(x-y)}\tilde{G}(k)$) leads us to:
\begin{align}\label{gf1}
(\mathbf{M}_0(k)+\mathbf{M}_1(h,\t))\tilde{G}=\i.
\end{align}
in which we can have $\hat{G}=\hat{G}_0+\hat{G}_1$, as a perturbative solution.
With a simple substitution, Eq.(\ref{gf1}) becomes:
\begin{align}\label{fgf}
\tilde{G}_1=-\frac{1}{\mathbf{M}_0(k)}\mathbf{M}_1(h,\t)\tilde{G}_0,
\end{align}
where $\tilde{G}_0=\frac{\i}{k\sl-m\22}$ corresponds to free fields. 
By an analysis similar to that which lead to the above result, we see that:
\begin{align}\label{fgff}
\tilde{G}_1=2\frac{h-\frac{1}{2}\t^{0i}\dot{h}k_i}{k^2-m^2}k_1k_2\tilde{G}_0+\i\frac{\mathbf{M}_1(h,\t)}{k^2-m^2}.
\end{align}
Our calculation concludes that the Green's function is:
\begin{align}\label{fgfff}
\tilde{G}=
\tilde{G}_0
+2\frac{h-\frac{1}{2}\t^{0i}\dot{h}k_i}{k^2-m^2}k_1k_2\tilde{G}_0
+\i\frac{\mathbf{M}_1(h,\t)}{k^2-m^2}.
\end{align}
We see that the probability density of finding a constitute of the particle is effectively time dependent by Eq.(\ref{fgfff}). 
Also the obtained result indicates the particles and antiparticles have pure states which are independent of one another.

\section{Discussion}
We consider the quantum dynamics of spinorial fields in another Minkowski non commutative space time.
We also calculate vierbein fields in four-dimensional space time in the presence of a gravitational background.
Based on DeWitt's method, an action will be derived for a perturbed Dirac theorem by employing vierbein fields.
By introducing non commutative coordinates and by making the replacement of ordinary product with the star product everywhere, 
we construct the temporal non commutative version of Dirac theorem. 
We can derive the first corrections of the solution and Green's function in terms of Riemann's curvature tensor.
The obtained result indicates that the new state of the particles are pure state which are independent of one another.
We conclude that the matter fields get the different energies and masses related to their frequency.     
The separation of the solutions is realized by identifying frequency.

\section{Acknowledgements}
The author thanks the Shahrekord University for supporting of this work through a research grant.
\newline

\end{document}